\documentclass[twocolumn,twoside,pra,superscriptaddress]{revtex4}
\usepackage[utf8]{inputenc}
\usepackage{amsmath}
\usepackage{amssymb}
\usepackage{amsthm}
\usepackage{comment}
\usepackage{latexsym,revsymb}
\usepackage{hyperref}

\usepackage{mathtools}

\newcommand{\av}[1]{\langle #1 \rangle}
\newcommand{\de}{d_{\mathrm{eff}}}
\newcommand{\f}[2]{\textstyle{\frac{#1}{#2}}}
\newcommand{\tr}[1]{\mathrm{tr}\! \left[#1\right]}

\newcommand{\abs}[1]{\left| #1 \right|} 

\usepackage{color}
\newcommand{\tcf}[1]{#1}

\makeatletter
\newtheorem*{rep@theorem}{\rep@title}
\newcommand{\newreptheorem}[2]{%
\newenvironment{rep#1}[1]{%
 \def\rep@title{#2 \ref{##1}}%
 \begin{rep@theorem}}%
 {\end{rep@theorem}}}
\makeatother

\newtheorem{theorem}{Theorem}
\newreptheorem{theorem}{Theorem}

\newreptheorem{corollary}{Corollary}
\newtheorem{lemma}[theorem]{Lemma}
\newreptheorem{lemma}{Lemma}
\newtheorem{Definition}{Definition}

\begin{document}
\title{Thermalization and Return to Equilibrium 
on Finite Quantum Lattice Systems}

\author{Terry Farrelly}
\affiliation{Institut f{\"u}r Theoretische Physik, Leibniz Universit{\"a}t, 30167 Hannover, Germany}
 \author{Fernando G.S.L. Brand\~ao}
\affiliation{IQIM, California Institute of Technology, Pasadena CA 91125, USA}
\author{Marcus Cramer}
\affiliation{Institut f{\"u}r Theoretische Physik, Leibniz Universit{\"a}t, 30167 Hannover, Germany}
%

\begin{abstract}
 Thermal states are the bedrock of statistical physics.  Nevertheless, when and how they actually arise in closed quantum systems is not fully understood.  We consider this question for systems with local Hamiltonians on finite quantum lattices.  In a first step, we show that states with exponentially decaying correlations equilibrate after a quantum quench. Then we show that the equilibrium state is locally equivalent to a thermal state, provided that the free energy of the equilibrium state is sufficiently small and the thermal state has exponentially decaying correlations.  As an application, we look at a related important question:\ When are thermal states stable against noise?  In other words, if we locally disturb a closed quantum system in a thermal state, will it return to thermal equilibrium?  We rigorously show that this occurs when the correlations in the thermal state are exponentially decaying.  All our results come with finite-size bounds, which are crucial for the growing field of quantum thermodynamics and other physical applications.
\end{abstract}

\maketitle
\section{Introduction}
To understand the strengths and limitations of statistical physics, it makes sense to derive it from physical principles, without ad hoc assumptions.  Along these lines, over the past twenty years, ideas from quantum information have \tcf{given} new insights into the foundations of statistical physics \cite{EFG14,GHRRS15,GE15}.  In particular, some progress \tcf{was} made towards understanding how and when thermalization occurs \cite{Reimann10,RGE11,MAMW13}.  A large class of states of systems with weak intensive interactions (e.g.,\ one dimensional systems) \tcf{were shown to thermalize \cite{RGE11}}.  In \cite{MAMW13} thermalization was \tcf{proved}, also for a large class of states, in the thermodynamic limit.  (We will compare the results of \cite{MAMW13} to ours in detail below.)  More recently, the equivalence of the microcanonical and canonical ensemble (i.e., thermal state) was proved for finite quantum lattice systems, when correlations in the thermal state decay sufficiently quickly \cite{BC15} (see also \cite{MAMW13,Tasaki16}).

Here, we prove thermalization results for closed quantum systems in two parts.  First, we build upon previous equilibration results (\tcf{e.g., Refs \cite{Tasaki98,LPSW09}}).  A requirement for equilibration is that the effective dimension, defined 
below, is large.  While there are physical arguments for this \tcf{in some cases} \cite{Reimann08} \tcf{(}and it is true for most states drawn from the Haar measure on large subspaces \cite{LPSW09}\tcf{)}, there are no techniques to decide whether a given initial state will equilibrate under a given Hamiltonian. Here, we prove that a large effective dimension is guaranteed for local Hamiltonian systems if the correlations in the initial state decay sufficiently quickly and the energy variance is sufficiently large. The latter is known for thermal states with intensive specific heat capacity and may, for large classes of states, be computed straightforwardly. The second part of thermalization is to show that the equilibrium state is locally indistinguishable from a thermal state. We prove that this occurs if the correlations in the corresponding thermal state decay sufficiently quickly and the relative entropy difference between the equilibrium state and the corresponding thermal state is sufficiently small.

As an application, we answer the following question. Given a closed quantum system that is initially in a thermal state, and suppose we locally quench or disturb it, will it re-equilibrate to a thermal state?  
Understanding when thermal states are robust against local external noise is important, e.g., in decohering quantum simulations implemented in optical lattice systems (see, e.g., Ref.\ \cite{SPTD15} and references therein), where noise can be caused by the absorption and re-emission of a photon. \tcf{Q}uestions of re-equilibration have a long tradition, and
return to equilibrium was shown for infinite lattice systems in the seventies \cite{Rob73,HR86}, by making transport assumptions:\ On an infinite lattice, information may leave a region and never return, which is not true for finite systems. This fundamental difference highlights the importance of finite-size considerations.
 We discuss the connection to results on infinite lattices further in appendix \ref{sec:Robinson}.
 
 Return to equilibrium \tcf{was} also shown for finite quantum systems coupled to infinite reservoirs after a coupling has been turned on \cite{BFS00,JPPP15}, and a rough argument for stability of thermal states was given recently in terms of energy probability distributions in \cite{HF16}.  
Here, we prove that a system in a thermal state, after being locally disturbed, re-equilibrates to a thermal state provided correlations decay sufficiently quickly.  
In contrast to infinite lattice systems, our results give finite-size estimates and our methods and assumptions are entirely different.  We emphasize that finite-size bounds are crucial for physical applications, particularly those in quantum thermodynamics \cite{GHRRS15}, where thermal states are usually considered to be free resources. \tcf{Understanding} to what extent thermalization occurs will also affect protocols for extracting work using small quantum thermal machines. 

\section{Setting and notation}
We consider a $d$-dimensional hypercubic lattice with $N=n^d$ sites. We suppose that each site has a $d_{\mathrm{loc}}$-dimensional quantum system, e.g., a spin.  We let $H$ \tcf{denote the ($k$-local) Hamiltonian}, i.e., it \tcf{has} the form $H=\sum_ih_i$, where $h_i$ acts only on lattice sites that are no more than $k$ sites from $i$, i.e., only on sites $j$ with $\text{dist}(i,j)\le k$. We further assume the $h_i$ are bounded in operator norm
and use units with $\|h_i\|\le 1$ and $\hbar=k_B=1$. We write $\rho(t)=e^{-\mathrm{i}Ht}\rho\, \mathrm{e}^{\mathrm{i}Ht}$ for the state at time $t$ and denote the time-average state by
\begin{equation}
\langle\rho\rangle=\lim_{T\rightarrow\infty}\frac{1}{T}\int_0^T\!\mathrm{d}t\,\rho(t).
\end{equation}
We let $\sigma^2$ denote the energy variance of a state $\rho$ with respect to a Hamiltonian $H$, $\sigma^2=\tr{\rho H^2}-\mathrm{tr}^2[\rho H]$. 
We will be interested in subsystems, $S$, of the whole lattice and denote the rest by $B$ -- the {\em bath} or {\em environment}. We denote their Hilbert space dimensions by $d_S=d_{\mathrm{loc}}^{|S|}$ and $d_B=d_{\mathrm{loc}}^{|B|}$. Given a state of the whole system $\rho$, we write $\rho_S=\mathrm{tr}_B[\rho]$ and $\rho_B=\mathrm{tr}_S[\rho]$ for the reduced states o\tcf{f} the subsystem and environment, respectively.

To discuss whether two states are close, we consider what one can measure in practice.  Mostly, we will consider the local distinguishability of two states, $\rho$ and $\tau$, given by
$\|\rho-\tau\|_S:=\|\rho_S-\tau_S\|_{\mathrm{tr}},$
where $\|\cdot \|_{\mathrm{tr}}$ is the trace distance.  \tcf{O}ur results extend naturally to coarse-grained observables.  An example of \tcf{which} could be the magnetization of spins on a large region or even the whole lattice.
We may write \tcf{such an} observable as $M=\frac{1}{m}\sum_{i=1}^m M_{S_i}$, where $S_i$ are non-overlapping subsystems and $M_{S_i}$ acts only on subsystem $S_i$.  For example, one could take the magnetization per spin $M=\frac{1}{N}\sum_i \sigma_z^i$. Then local indistinguishability implies that expectation values of such observables are close: Assuming $\|M_{S_i}\|\le C$, we have
\begin{equation}
 \abs{\tr{\rho M}-\tr{\sigma M}} \leq  C\cdot\mathbb{E}_{S_i}\|\rho-\sigma\|_{S_i},
\end{equation}
where $\mathbb{E}_{S_i}$ denotes the average over subsystems $S_i$. Thus, we cover many physically realistic measurement\tcf{s}.

Throughout, we will often consider states with exponentially decaying correlations.  This is guaranteed for, e.g., thermal states above a critical temperature \cite{KGKRE14} and ground states of gapped $k$-local Hamiltonians \cite{HK06}.  We define exponentially decaying correlations \tcf{as follows}.
\begin{Definition}
\label{exp dec cor}
 A state $\rho$ has exponentially decaying correlations if there is a correlation length $\xi>0$ and a $K\geq 0$ (both independent of the system size $N$), such that, for any two lattice regions $X$ and $Y$, one has
 \begin{equation}
 \nonumber
  \max_{\substack{\text{supp}[P]\subset X\\ \text{supp}[Q]\subset Y\\ \|P\|,\|Q\|=1}}\abs{\frac{\tr{\rho P Q}- \tr{\rho P}\tr{\rho Q}}{|X||Y|}}\leq Ke^{-\mathrm{dist}(X,Y)/\xi}.
 \end{equation}
\end{Definition}
\noindent
Here, the distance between the regions $X$ and $Y$ is
$\mathrm{dist}(X,Y)=\min_{i\in X,j\in Y}\mathrm{dist}(i,j)$,
where $\mathrm{dist}(\cdot,\cdot)$ is some metric on the lattice.

\section{Equilibration}\label{sec:equilibration}
Due to recurrences, a closed finite system will never truly equilibrate, not even locally. Hence, for finite systems one asks a different question \cite{Reimann08,LPSW09}: Does a system spend most of its time close to some fixed state? If we denote the fixed state by $\tau$, this means that 
\begin{equation}
\label{local equilibration}
D_S(\tau):=
\lim_{T\rightarrow\infty}\frac{1}{T}\int_0^T\!\!\mathrm{d}t\,\|\rho(t)-\tau\|_S
\end{equation}
is small, i.e., for the majority of times, $\rho(t)$ and $\tau$ are locally indistinguishable.  The most natural \tcf{case} is equilibration to the time-average state.  For this case, it was proved that \cite{LPSW09,SF12}
\begin{equation}\label{eq:LPSW}
D_S(\av{\rho})\leq \frac{1}{2}\sqrt{\frac{D_G d_S^2}{\de}},
\end{equation}
 where $d_S$ denotes the Hilbert space dimension of the subsystem $S$.
Here, $D_G$ is the degeneracy of the most degenerate \emph{energy gap} \footnote{Labelling non-zero energy gaps by $G_{\alpha}=G_{ij}=E_i-E_j$, one defines
$D_G=\max_{\alpha}|\{G_{\beta}|G_{\beta}=G_{\alpha}\}|$.}, i.e., $D_G=1$ if \tcf{there are} no degenerate energy gaps.
Typically, one expects $D_G$ to be small.  Actually, the existence of degenerate energy gaps is a measure zero constraint on the Hamiltonian.  Also in equation (\ref{eq:LPSW}) is $\de$, known as the effective dimension, defined by
\begin{equation}
 \frac{1}{\de}=\sum_k \mathrm{tr}^2[P_k\rho],
\end{equation}
where $P_{k}$ is the energy projector corresponding to energy $E_k$. If the energies are non-degenerate, then $\tcf{1/\de=} \tr{\langle\rho\rangle^2}$, i.e., the purity of the equilibrium state. Equation~(\ref{eq:LPSW}) implies that equilibration occurs when $\de$ is large.  The fraction of times \tcf{when} $\|\rho(t)-\av{\rho}\|_S\geq \delta$ is at most $(d_S\sqrt{D_G})/(2\delta\sqrt{\de})$, \tcf{via} Markov's inequality.  \tcf{Equation~(\ref{eq:LPSW})} is quite powerful:\ It holds for any decomposition of the total system into a subsystem $S$ and bath $B$.  This division need not correspond to a spatially localized subsystem.  For example, one could \tcf{consider} multi-point correlation functions over arbitrary distances.

In Ref.~\cite{Reimann08} it was argued on physical grounds that we should expect $\de$ to be exponentially large in the system size.  The argument relied on the exponentially increasing density of energy levels for generic physical systems.
\tcf{Also, if the initial state is thermal, then $1/\de\leq \tr{\rho(0)^2}\leq e^{\beta F}$, where $F=-1/\beta \ln(Z)$ is the free energy.}

However, one \tcf{cannot use} these arguments in all \tcf{physically interesting} situations:\ E.g., \tcf{for} a local (or global) quench.  \tcf{The free energy argument above is only useful for a highly mixed initial state, in contrast to initial pure or low temperature states.}  And there are \tcf{simple examples} where the initial state will not have an effective dimension that is exponentially large.  \tcf{For example, take the ground state of $H=-b\sum_i\sigma_x^i$.  After quenching to $H=-J\sum_i\sigma_z^i\sigma_z^{i+1}-h\sum_i\sigma_z^i$, the effective dimension is at most $O(N^2)$ \footnote{This is because the number of energy levels is $O(N^2)$, so $\de\leq O(N^2)$.  This follows because the effective dimension is bounded by the number of energy levels.}.}  Furthermore, calculating the effective dimension means computing the overlaps of the state with energy eigenvectors, which is as hard as diagonalizing the Hamiltonian.

So we need concrete lower bounds on the effective dimension.  Here we prove such a bound.
\begin{lemma}\label{lem:4}
Suppose the initial state $\rho$ $($or its time average $\av{\rho})$ has exponentially decaying correlations as in Def.~\ref{exp dec cor}.  Let the system evolve according to a bounded $k$-local Hamiltonian, and let $\rho$ have energy variance $\sigma^2$ with respect to this Hamiltonian. Then there is a constant $C$ independent of $N$ such that
 \begin{equation}
  \frac{1}{\de}\leq  C\frac{\ln^{2d}(N)}{s^3\sqrt{N}},
 \end{equation}
 where $s=\sigma/\sqrt{N}$.
\end{lemma}
This is proved in appendix \ref{app:Berry-Esseen} via a \tcf{quantum} Berry-Esseen theorem \cite{BC15,CBG16}. By assuming exponential decay of correlations, $s$ is upper bounded independent\tcf{ly} of $N$. Often, it is also lower bounded, e.g., for thermal states with intensive specific heat capacity $c(\beta)$, which is given at inverse temperature $\beta$ by $c(\beta)=\beta^2\sigma^2/N=\beta^2s^2$. 
Furthermore, for \tcf{many} states (e.g., product states or matrix product states) it is straightforward to compute $\sigma^2$ so the question of equilibration may be answered directly, without \tcf{knowing} the overlap of the initial state with the energy eigenstates. \tcf{T}hat $\sigma$ needs to be sufficiently large is reasonable: If the initial state is not sufficiently spread over many eigenstates\tcf{,} one cannot expect equilibration.

\tcf{We may use Lemma~\ref{lem:4} with Eq.~(\ref{eq:LPSW}) to show that equilibration occurs.  One situation where this is interesting is a} quench. If the initial state is a ground state of some Hamiltonian with exponentially decaying correlations (e.g., the ground state of a gapped $k$-local Hamiltonian), then after quenching to any other local Hamiltonian, equilibration will occur provided the energy variance $\sigma^2$ (with respect to the post-quench Hamiltonian) is sufficiently large.
\tcf{Note that we need the total system to be quite large, with $\sqrt{N}\gg d^2_S$.  Lemma~\ref{lem:4} can also be applied to equilibration in the settings of \cite{Reimann08,Short10}.}

\tcf{An immediate application of Lemma~\ref{lem:4} is to models studied in \cite{BCH11}.  In the examples where weak thermalization and no thermalization are observed (see Figures $1$ and $2$ in \cite{BCH11}), it is not clear whether equilibration will occur at all.  However, in both cases $\sigma$ is $O(\sqrt{N})$, and we can apply Lemma~\ref{lem:4} to lower bound the effective dimension.  This guarantees that equilibration will occur for a finite model, provided there are not too many degenerate energy gaps, which is reasonable as the models are far from integrable.}

Finally, we should mention that \tcf{little is known about equilibration timescales}.  Rigorous \tcf{bounds for} general \tcf{systems are} extremely large \cite{SF12,GHT13,MGLFS14} (these \tcf{often} involve $d_{\mathrm{eff}}$\tcf{,} so our results \tcf{apply}).  For some quadratic models, the timescale is much shorter \cite{CDEO08,TCF15}.  \tcf{Shorter timescales were also found for} random Hamiltonians, states or measurements \cite{MGLFS14,VZ12,Cramer12,BCHHKM12,UWE12,MRA13,Reimann16}.  Still, examples exist of reasonable translationally invariant models with extremely long equilibration timescales \cite{SSM15}.

\section{Thermalization}\label{sec:Thermalization}
In the previous section, we \tcf{saw} that equilibration occurs with great generality to the time-average state $\langle\rho\rangle$.  But that is only part of thermalization.  The second part is \tcf{that} the time-average state \tcf{must be} close to a thermal state. We thus need a practical way to decide whether, locally, $\langle\rho\rangle$ is close to a thermal state. The following Lemma (see appendix \ref{app:therm} for a more quantitative version), recently obtained in Ref.~\cite{BC15}, aids this by relating the local trace-norm distance of two states to their relative entropy \tcf{difference}.

\begin{lemma}\label{lem:BC main text} Let $\sigma$ be a state with exponentially decaying correlations as in Def.~\ref{exp dec cor}. Let $0<\alpha<\frac{1}{d+2}$ and let $l\in\mathbb{N}$, $l^d\in o(n^{\frac{1-\alpha}{d+1}})$. Let $\tau$ be a state. If
\begin{equation}\label{eq:BC}
\begin{split}
S(\tau\|\sigma)\in o\bigl(N^{\frac{1-(2+d)\alpha}{d+1}}\bigr),
\end{split}
 \end{equation}
then there is a constant $C$, independent of $N$, such that the average local trace distance between $\sigma$ and $\tau$ is bounded\tcf{:}
 \begin{equation}
 \label{eq:average bound}
  \mathbb{E}_{S\in\mathcal{S}_l}\|\sigma-\tau\|_S \leq \frac{C}{N^{\alpha/2}},
 \end{equation}
 where $\mathbb{E}_{S\in\mathcal{S}_l}$ denotes the average over all hypercubes on the lattice with length of side $l$.
\end{lemma}
Note that even if the relative entropy difference between the two states increases with system size (as in Eq.~\eqref{eq:BC}), the two states are locally close (on average over all cubic subsystems of size $l^d$). Also, maybe surprisingly, the size of the subsystem need not be fixed but may increase as a power law in $N$. The bound in Eq.~\eqref{eq:average bound} tells us that (if $N$ is sufficiently large)\tcf{,} for the vast majority of subsystems $S\in \mathcal{S}_l$\tcf{,} the states $\sigma_S$ and $\tau_S$ are close.
\tcf{For} course-grained observables\tcf{,} as discussed below, one finds, e.g., for the magnetization per spin $M=\frac{1}{N}\sum_i\sigma_i^z$ (\tcf{with} $l=1$)
\begin{equation}
\begin{split}
 \abs{\tr{\sigma M}-\tr{\tau M}}  \leq 
 \mathbb{E}_{S\in\mathcal{S}_l}\|\sigma-\tau\|_S,
 \end{split}
\end{equation}
\tcf{so} the bound in Eq.~\eqref{eq:average bound} directly gives a bound on the difference of expectation values in $\sigma$ and $\tau$.
If both states are translationally invariant, the average is obsolete, and one has $\|\sigma-\tau\|_S \leq CN^{-\alpha/2}$ for {\em all} cubic $S$ of size $l^d$.

Let us move on to thermalization, i.e., we \tcf{wish} to show that $D_S(\rho_{\beta})$ is small and hence that for \tcf{most} times $\rho(t)$ is locally close to the thermal state 
 $\rho_{\beta}=\mathrm{e}^{-\beta H}/Z$, $Z=\mathrm{tr}[\mathrm{e}^{-\beta H}]$. We do \tcf{this} by combining Eq.~\eqref{eq:LPSW} with Lemma~\ref{lem:4} and \ref{lem:BC main text}:
Let the initial state $\rho$ (or $\langle\rho\rangle$) have exponentially decaying correlations, evolve \tcf{via} a bounded $k$-local Hamiltonian $H$, and have energy variance $\sigma^2$.
Fix $l\in\mathbb{N}$ and $\alpha\in\mathbb{R}$, $0<\alpha<\frac{1}{d+2}$. Let the thermal state $\rho_\beta$ have exponentially decaying correlations and suppose
\begin{equation}
\label{eq:rel ent}
S(\langle\rho\rangle\|\rho_{\beta})\in o\bigl(N^{\frac{1-(2+d)\alpha}{d+1}}\bigr).
\end{equation}
Then there is a constant $C$ independent of $N$ such that (see appendix \ref{app:therm} for details)
\begin{equation}
\label{thermalization corollary}
\begin{split}
\mathbb{E}_{S\in\mathcal{S}_l}D_S(\rho_{\beta})
&\le C\left(\sqrt{\frac{D_G}{s^3N^{\frac{d}{2d+4}}}}+1\right)\frac{1}{N^{\alpha/2}}.
\end{split}
 \end{equation}
If $\rho_\beta$ and $\langle\rho\rangle$ are translationally invariant then $D_S(\rho_{\beta})$ is upper bounded by the right-hand side for {\em all} $S\in\mathcal{S}_l$, i.e., thermalization \tcf{occurs} on {\em every} cubic subsystem of size $l^d$. This is true, e.g., when the Hamiltonian is translationally invariant \tcf{with} no degenerate energies. Without requiring translational invariance or making some other transport assumption, we cannot guarantee that every subsystem thermalizes.  This is reasonable:\ We could consider models where \tcf{a few} small subsystems retain memory of their initial state.

In fact, we could replace the \tcf{assumption} that $\langle\rho\rangle$ be translationally invariant by \tcf{assuming} transport in the following sense.  Suppose that, in terms of the time-average state, one cannot tell where a localized disturbance of the initial state had occurred.  In other words, let $\Phi_i$ denote a local quantum channel on some region centred on $i$.  Then we demand that $\|\av{\Phi_i(\rho)}-\av{\Phi_j(\rho)} \|_S\leq \epsilon\ll 1$ for any $i$, $j$ and some small region $S$.  \tcf{Therefore,} locally the equilibrium state is indistinguishable from $\av{\f{1}{N}\sum_{i}^{N}\Phi_i(\rho)}$, which is translationally invariant.  \tcf{So} the thermalization result Eq.\ (\ref{thermalization corollary}) holds for the \emph{individual} subsystem $S$ with an extra $\epsilon$ on the right hand side.  \tcf{This follows from} the triangle inequality.  We discuss transport assumptions further in appendix \ref{sec:Robinson}.

It is important to compare Eq.~(\ref{thermalization corollary}) to the results of \cite{MAMW13}, which proved that thermalization occurs in the thermodynamic limit, with a comparable condition	 on the time-average state.  Here, we prove thermalization for the important case of finite systems\tcf{,} and \tcf{we can} give finite-size estimates.  Furthermore, in \cite{MAMW13} the thermal state \tcf{must} correspond to a unique phase.  Instead, we assume that the thermal state has exponentially decaying correlations. This is always satisfied for $d=1$ \cite{Araki69} and, for $d>1$, if the temperature is above a critical temperature \cite{KGKRE14}.

 Finally, we note that the free energy of a state $\rho$ at inverse temperature $\beta$ is given by $F(\rho)=\tr{H\rho}-S(\rho)/\beta$, so
$S(\langle\rho\rangle\|\rho_{\beta})=\beta\left(F(\langle\rho\rangle)-F(\rho_{\beta})\right)$.
Thus, whenever the free energy of $\langle\rho\rangle$ is sufficiently small, $\langle\rho\rangle$ and $\rho_{\beta}$ are locally close.

\section{The stability of thermal states}
We can apply these results to some interesting examples.  We will focus on the translationally invariant setting, i.e., we will assume that the time-average state $\langle\rho\rangle$
and the thermal state $\rho_\beta$ are translationally invariant. This is true, e.g., when the Hamiltonian is translationally invariant and has no degenerate energies. As discussed above, translational invariance guarantees transport, without which we can not expect all subsystems to thermalize.

For the first example, suppose we have a system that was in a thermal state $\rho_{\beta}$, but was affected by a local process or some localized noise.  We can model this by \tcf{applying} a local quantum channel \footnote{We may take $\Phi(\rho_\beta)$ as the initial state, where
$\Phi(\rho_\beta)=\sum_iK_i^\dagger\rho_\beta K_i$, with $\sum_iK_iK_i^\dagger=\mathbb{I},$
and the $K_i$ act only locally.}.
  We \tcf{now} see that the system locally returns to thermal equilibrium provided $\rho_{\beta}$ had exponentially decaying correlations.

\begin{theorem}\label{lem:2}
Let $H$ be a bounded $k$-local Hamiltonian. Let $\rho_{\beta}$ be a translationally-invariant thermal state with exponentially decaying correlations as in Def.~\ref{exp dec cor} and energy variance $\sigma^2$. Suppose $\Phi$ is a quantum channel acting non trivially only on a cubic subsystem of fixed size. Fix $l\in\mathbb{N}$. Let $\rho=\Phi(\rho_\beta)$ evolve under $H$, and let $\langle\rho\rangle$ be translationally invariant.
Then the system locally re-thermalizes:\ There is a constant $C$ independent of $N$ such that
 \begin{equation}
 \label{thing}
\begin{split}
D_S(\rho_{\beta})
&\le C\left(\sqrt{\frac{D_G}{s^3N^{\frac{d}{2d+4}}}}+1\right)\frac{1}{N^{\frac{1}{2d+5}}}
\end{split}
 \end{equation}
 for all cubic subsystems $S$ of size $l^d$, i.e., the system re-thermalizes on any cubic subsystem of fixed size.  In particular, this is true for the subsystem on which the channel $\Phi$ acted.  See appendix \ref{app:corollaries} for the proof.
\end{theorem}

\tcf{As a second example, consider} a system in thermal equilibrium\tcf{.}  \tcf{H}ow much may the Hamiltonian change \tcf{such that} the system \tcf{still} equilibrate\tcf{s} to a thermal state?
The following theorem gives a rigorous answer.
\begin{theorem}\label{lem:3} Let $H_0$ be a Hamiltonian and $\rho=e^{-\beta H_0}/Z_0$ be the system's initial state, which we assume to have exponentially decaying correlations.  Suppose that this state evolves under a bounded $k$-local Hamiltonian $H$ and has energy variance $\sigma^2$ with respect to $H$. Let $\rho_\beta=\mathrm{e}^{-\beta H}/Z$, the thermal state corresponding to $H$, be translationally invariant and have exponentially decaying correlations. Let $\langle\rho\rangle$ be translationally invariant and $0<\alpha<\frac{1}{d+2}$. If
$\|H-H_0\|\in o\bigl(N^{\frac{1-(2+d)\alpha}{d+1}}\bigr)$
then there is a constant $C$ independent of $N$ such that
\begin{equation}
\begin{split}
D_S(\rho_{\beta})
&\le C\left(\sqrt{\frac{D_G}{s^3N^{\frac{d}{2d+4}}}}+1\right)\frac{1}{N^{\alpha/2}}
\end{split}
 \end{equation}
 for all cubic subsystems $S$ of size $l^d$.  See appendix \ref{app:corollaries} for a proof.  
\end{theorem}
This implies that we can quench from Hamiltonian $H_0$ to Hamiltonian $H$, and we get local thermalization to $\rho_\beta$ for any cubic subsystem of fixed size $l^d$ provided the Hamiltonians are not too different. Maybe surprisingly, we are not restricted to local quenches: The difference between the Hamiltonians may grow as a power law in the system size $N$.

\section{Discussion}
We have seen that, after locally perturbing a quantum system in a thermal state (with exponentially decaying correlations), the system equilibrate\tcf{s} to a state \tcf{in}distinguishable from \tcf{a} thermal state on small subsystems of fixed size.  This \tcf{may not be} true if there are long-range correlations in the initial state.  Also, notice that one can easily construct counterexamples where an individual small subsystem will not return to thermal equilibrium after being perturbed \emph{without} some form of transport assumption.

In \cite{Rob73} there are infinite lattice analogues of our findings (which are for finite systems).  Infinite lattices are a\tcf{n entirely} different setting because information can leave a subsystem and never return. \tcf{Nevertheless, one may draw inspiration from \cite{Rob73} and try to} generalize our work:\ E.g., it may be possible to go beyond thermal states and show return to equilibrium of more general equilibrium states.  \tcf{W}e discuss how one may approach this problem further in appendix \ref{sec:Robinson}.

\acknowledgments
The authors are grateful to Tobias Osborne and David Reeb for helpful discussions.  This work was supported by the ERC grants QFTCMPS and SIQS, by the cluster of excellence EXC201 Quantum Engineering and Space-Time Research, and by the DFG through SFB 1227 (DQ-mat).

\bibliographystyle{unsrt}

\appendix

\section{A lower bound for the effective dimension}
\label{app:Berry-Esseen}
To lower bound the effective dimension for local Hamiltonian models, we will use a theorem from \cite{BC15,CBG16} as a stepping stone.  This is a quantum version of the Berry-Esseen theorem.  The Berry-Esseen theorem is a more powerful statement than the central limit theorem, as it gives the rate of convergence of a distribution to a Gaussian. Let $H=\sum_\nu E_\nu |\nu\rangle\langle \nu|$ be $k$-local, i.e., let $H$ be of the form $H=\sum_ih_i$ with $h_i$ acting only on sites $j$ with $\text{dist}(i,j)\le k$. Let $\rho$ have exponentially decaying correlations as in Def.~1. Then, by Lemma 8 of Ref.~\cite{BC15} 
\begin{equation}
\label{QBE}
\begin{split}
\Delta&:=\sup_{x}\bigl|F(x)-G(x)\bigr|\\
&\le  C_d\left((k+\xi)\left(\frac{\ln(K)}{\ln(N)}+3\right)\right)^{2d}\\
&\hspace{2cm}\times\left(1+\frac{s^2}{\ln(N)}\right)\frac{\ln^{2d}(N)}{s^3\sqrt{N}},
\end{split}
\end{equation}
where $C_d$ depends only on the lattice dimension $d$ and we recall that $s=\sigma/\sqrt{N}$. 
Here, $F$ and $G$ are the cumulative distribution functions
\begin{equation}
\begin{split}
F(x)&=\sum_{\nu:E_\nu\le x}\langle \nu|\rho|\nu\rangle,\\
G(x)&=\frac{1}{\sqrt{2\pi\sigma^2}}\int_{-\infty}^{x}\mathrm{d}y\,e^{-\frac{(y-\tr{\rho H})^2}{2\sigma^2}}.
\end{split}
\end{equation}
We may simplify the upper bound on $\Delta$ by noting that $s$ is upper-bounded independent of $N$: Write $v_{k,d}=\max_i|\mathrm{supp}[h_i]|$. Then
\begin{equation}
\begin{split}
s^2&=\frac{1}{N}\sum_{i,j}\langle h_ih_j\rangle-\langle h_i\rangle\langle h_j\rangle\\
&\le Kv_{k,d}^2 \frac{1}{N}\sum_{i,j}\mathrm{e}^{-\text{dist}(\mathrm{supp}[h_i],\mathrm{supp}[h_j])/\xi}\\
&\le Kv_{k,d}^2\,\mathrm{e}^{2k/\xi} \sum_{l=0}^\infty\mathrm{e}^{-l/\xi}b_{d,l},
\end{split}
\end{equation}
where $b_{d,l}=\max_i|\{j\,|\, \text{dist}(i,j)=l\}|$ is the maximum surface area of a ball of radius $l$ centred at $i$. Hence, we have that there is a constant $C$ independent of $N$ such that 
\begin{equation}
\begin{split}
\Delta&\le  C\frac{\ln^{2d}(N)}{s^3\sqrt{N}}.
\end{split}
\end{equation}

To apply the above theorem, we note that for any $\epsilon>0$
\begin{equation}
\begin{split}
\frac{1}{d_{\mathrm{eff}}}&=\sum_{\nu}\mathrm{tr}^2[\rho P_\nu]\\
&\leq \max_{\nu}\mathrm{tr}[\rho P_\nu]\\
& \le\max_x \left[F( x)-F( x-\epsilon)\right].
\end{split}
\end{equation}
Applying the bound in Eq.~\eqref{QBE} and the mean value theorem we thus find
\begin{equation}
\begin{split}
\frac{1}{d_{\mathrm{eff}}}&\le 2\Delta+\max_x \left[G(x)-G(x-\epsilon)\right]\\
&\le 2\Delta+\epsilon.
\end{split}
\end{equation}
As $\epsilon$ was arbitrary we thus have that
\begin{equation}
\label{eff dim bound}
\frac{1}{d_{\mathrm{eff}}}\le 2\Delta.
\end{equation}
 
If we instead assume that $\langle\rho\rangle$ has exponentially decaying correlations, we use 
\begin{equation}
\tr{\rho P_\nu}= \tr{\langle\rho\rangle P_\nu}
\end{equation}
to arrive at the same bound on $d_{\mathrm{eff}}$. Note also that $\tr{\rho H}=\tr{\langle\rho\rangle H}$ and $\tr{\rho H^2}=\tr{\langle\rho\rangle H^2}$.

\section{Lemma \ref{lem:BC main text} and proof of Eq.~(\ref{thermalization corollary})}
\label{app:therm}
We rely on Proposition $2$ of \cite{BC15}, which we state in a slightly simplified version:


\begin{lemma}\label{lem:BC} Let $\sigma$ be a state with exponentially decaying correlations as in Def.~1. Let $\alpha>0$ and let $N$ and $l\in\mathbb{N}$ such that $l\le \frac{n+1}{2}$ and
\begin{equation}
\label{condition local equivalence}
3N^{\alpha}+\frac{2\xi\ln(d_{\mathrm{loc}})+3}{\xi\ln(2)}l^d+\log(N^{\frac{\ln(K)}{\ln(N)}+3})\le\frac{ 1}{4\xi^{\frac{d}{d+1}}}N^{\frac{1-\alpha}{d+1}}.
\end{equation}
Denote by $\mathcal{C}_l$ the set of all hypercubes on the lattice with length of side $l$. Let $\tau$ be a state. If
\begin{equation}
\begin{split}
S(\tau\|\sigma)\le\frac{1}{4\xi^{\frac{d}{d+1}}}N^{\frac{1-(2+d)\alpha}{d+1}}
\end{split}
 \end{equation}
then the average local trace distance between $\sigma$ and $\tau$ is bounded as
 \begin{equation}
   \mathbb{E}_{C\in\mathcal{C}_l}\|\sigma-\tau\|_C \leq \frac{7}{N^{\alpha/2}},
 \end{equation}
 where $\mathbb{E}_{C\in\mathcal{C}_l}$ denotes the average over all $C\in\mathcal{C}_l$.
\end{lemma}
We arrive at the Lemma in the main text by noting that Eq.~\eqref{condition local equivalence} is fulfilled for sufficiently large $N$ if $\alpha<\frac{1}{d+2}$
and $l^d\in o(N^{\frac{1-\alpha}{d+1}})$. Combining 
Eq.~\eqref{eq:LPSW} with Lemma~\ref{lem:4} and~\ref{lem:BC main text}, we find by the triangle inequality ($C$ denotes a constant independent of $N$, which may change from line to line)
\begin{equation}
\begin{split}
\mathbb{E}_{S\in\mathcal{S}_l}D_S(\rho_{\beta})
&\le \mathbb{E}_{S\in\mathcal{S}_l}D_S(\langle\rho\rangle)+\mathbb{E}_{S\in\mathcal{S}_l}\|\langle\rho\rangle-\rho_{\beta}\|_S\\
&\le \frac{1}{2}\sqrt{\frac{D_Gd_S^2}{\de}}+\frac{C}{N^{\alpha/2}}\\
&\le C\left(\sqrt{\frac{D_G \ln^{2d}(N)}{s^3\sqrt{N}}}+\frac{1}{N^{\alpha/2}}\right),
\end{split}
 \end{equation}
where
\begin{equation}
\begin{split}
\sqrt{\frac{\ln^{2d}(N)}{\sqrt{N}}}=\frac{\ln^d(N)}{N^{\frac{1}{2}(\frac{1}{d+2}-\alpha)}}\frac{1}{N^{\frac{d}{4(d+2)}}}\frac{1}{N^{\alpha/2}},
\end{split}
 \end{equation}
 which implies Eq.~\eqref{thermalization corollary} as $\frac{\ln^d(N)}{N^{\frac{1}{2}(\frac{1}{d+2}-\alpha)}}\rightarrow 0$ because $\alpha<\frac{1}{d+2}$.

\section{Proof of the theorems}
\label{app:corollaries}
To arrive at Theorem~\ref{lem:2} we need to (i) show that $\rho$ has exponentially decaying correlations, (ii) bound the relative entropy, and (iii) relate the energy variance of $\rho$ to that of $\rho_\beta$. 

We start with (i): If $X,Y$ are not in the subsystem that the channel acts on, then two-point correlations of operators $P$ and $Q$ with supports in $X$ and $Y$, respectively, decay exponentially as they do for $\rho_\beta$ as in Def.~\ref{exp dec cor}. Now denote by $A$ the subsystem the channel acts on and by $B$ the rest of the system. If $\mathrm{dist}(X,Y)>\mathrm{diam}(A)$ then either $X,Y\in B$ or one of the supports has overlap with $A$ and the other is contained in $B$. Let us consider the case $X\cap A\ne\emptyset$ and $Y\subset B$. Then, if $\rho=\Phi(\sigma)=\sum_iK_i^\dagger\sigma K_i$ with $K_i$ acting only on $A$ and $\sum_iK_iK_i^\dagger= \mathbb{I}$ and with adjoint $\Phi^*(\cdot)=\sum_iK_i\cdot K^\dagger_i$,
  \begin{equation}
  \begin{split}
&\tr{\rho P Q}- \tr{\rho P}\tr{\rho Q}=\tr{\rho P Q}- \tr{\rho P}\tr{\rho_\beta Q}\\
&\hspace{0.5cm}=\|\Phi^*(P)\|\frac{\tr{\rho_\beta \Phi^*(P) Q}- \tr{\rho_\beta \Phi^*(P)}\tr{\rho_\beta Q}}{\|\Phi^*(P)\|}\\
&\hspace{0.5cm}\le K|X\cup A||Y|\mathrm{e}^{-\mathrm{dist}(Z,Y)/\xi}\\
&\hspace{0.5cm}\le K|A||X| |Y|\mathrm{e}^{-\mathrm{dist}(Z,Y)/\xi}
 \end{split}
 \end{equation}
 as $\rho_\beta$ has exponentially decaying correlations and $\|\Phi^*(P)\|\le 1$ (via corollary $2.9$ in \cite{Paulsen02}). Here, $Z=\mathrm{supp}[\Phi^*(P)]\subset X\cup A$ such that $\mathrm{dist}(X,Y)\le \mathrm{dist}(Z,Y)+\mathrm{diam}(A)$. Hence, $\rho$ has exponentially decaying correlations with $K^\prime=|A|\mathrm{e}^{\mathrm{diam}(A)/\xi}K$. 

We now address (ii):\ Denote the subsystem on which $\Phi$ acts by $A$ and the rest of the system by $B$. Then $\rho=\Phi(\rho_\beta)$ and $\rho_\beta$ coincide on $B$. Writing $H=H_A+H_B$ with
$H_B$ acting exclusively on $B$ and $H_A$ collecting the remaining terms we have $\tr{H_B\Phi(\rho_\beta)}=\tr{H_B\rho_\beta}$ such that
  \begin{equation}
  \begin{split}
S(\langle\rho\rangle\|\rho_\beta)&=\beta\tr{H(\langle\rho\rangle-\rho_\beta)}+S(\rho_\beta)-S(\langle\rho\rangle)\\
&=\beta\tr{H(\Phi(\rho_\beta)-\rho_\beta)}+S(\rho_\beta)-S(\langle\rho\rangle)\\
&\le 2\beta\|H_A\|+S(\rho_\beta)-S(\langle\rho\rangle),
 \end{split}
 \end{equation} 
 where, as $|A|$ is independent of $N$ and $H$ is bounded and $k$-local, $\|H_A\|$ is bounded independent of $N$.
The entropy difference may be bounded by using $S(\langle\rho\rangle)\ge S(\rho)$ and the Araki--Lieb inequality $|S(\sigma_A)-S(\sigma_B)|\le S(\sigma)\le S(\sigma_A)+S(\sigma_B)$ \cite{AL70}, which holds for any state $\sigma$. We find
  \begin{equation}
  \begin{split}
&S(\rho_\beta)-S(\langle\rho\rangle)\\
&\hspace{1cm}\le S([\rho_\beta]_A)+S([\rho_\beta]_B)-|S(\rho_A)-S(\rho_B)|\\
&\hspace{1cm}\le S([\rho_\beta]_A)+S(\rho_A)\le 2|A|\ln(d_{\mathrm{loc}})
 \end{split}
 \end{equation} 
 as $[\rho_\beta]_B=\rho_B$. Hence, $S(\langle\rho\rangle\|\rho_\beta)$ is bounded from above by a constant independent of $N$. Thus, we may set $\alpha=\frac{1}{d+2.5}$ to find
 \begin{equation}
 \label{almost there}
\begin{split}
\mathbb{E}_{S\in\mathcal{S}_l}D_S(\rho_{\beta})
&\le C\left(\sqrt{\frac{D_G}{s_\rho^3N^{\frac{d}{2d+4}}}}+1\right)\frac{1}{N^{\frac{1}{2d+5}}},
\end{split}
 \end{equation}
 where $s_\rho^2=\sigma^2_\rho/N$ and $\sigma^2_\rho$ is the energy variance of the initial state $\rho=\Phi(\rho_\beta)$ with respect to $H$. Now,
  \begin{equation}
  \begin{split}
\sigma_\rho^2&=\sum_{i,j\in A}\left(\langle h_ih_j\rangle_{\rho}-\langle h_i\rangle_{\rho}\langle h_j\rangle_{\rho}\right)\\
&\hspace{1cm}+\sum_{i\in A,j\in B}\left(\langle h_ih_j\rangle_{\rho}-\langle h_i\rangle_{\rho}\langle h_j\rangle_{\rho}\right)\\
&\hspace{1cm}+\sum_{j\in A,i\in B}\left(\langle h_ih_j\rangle_{\rho}-\langle h_i\rangle_{\rho}\langle h_j\rangle_{\rho}\right)\\
&\hspace{1cm}+\sum_{i,j\in B}\left(\langle h_ih_j\rangle_{\rho}-\langle h_i\rangle_{\rho}\langle h_j\rangle_{\rho}\right)\\
 \end{split}
 \end{equation}  
 and similarly for $\rho_\beta$. As $\rho$ and $\rho_\beta$ coincide on $B$ and both states have exponentially decaying correlations and $|A|$ is upper-bounded independently of $N$, there is hence a constant $C$ independent of $N$ such that $|\sigma^2-\sigma_\rho^2|\le C$, i.e.,
 \begin{equation}
s^2_\rho\ge s^2\left(1-\frac{C}{Ns^2}\right).
 \end{equation}
If $s^{3/2}\le N^{-\frac{d}{4d+8}-\frac{1}{2d+5}}$ then Eq.~\eqref{thing} holds trivially, i.e., we may w.l.o.g. assume that $s^{3/2}\ge N^{-\frac{d}{4d+8}-\frac{1}{2d+5}}$. Then
$\frac{1}{s^{2}}\le N^{\frac{d}{3d+6}+\frac{4}{6d+15}}\le N^{\frac{1}{3}}$, i.e.,
 \begin{equation}
s^2_\rho\ge s^2\left(1-\frac{C}{N^{\frac{2}{3}}}\right)
 \end{equation} 
 such that for sufficiently large $N$ we have $s^2_\rho\ge s^2/2$, which, combined with Eq.~\eqref{almost there}, implies Eq.~\eqref{thing}.

To prove Theorem~\ref{lem:3} we need only bound the relative entropy. As above, we have
\begin{equation}
  \begin{split}
S(\langle\rho\rangle\|\rho_\beta) & =\beta\tr{H(\langle\rho\rangle-\rho_\beta)}+S(\rho_\beta)-S(\langle\rho\rangle)\\
&\le \beta\tr{H(\langle\rho\rangle-\rho_\beta)}+S(\rho_\beta)-S(\rho)\\
&= \beta\tr{(H-H_0)\rho}+\ln(Z)-\ln(Z_0)\\
&\le \beta\|H-H_0\|+\ln(Z)-\ln(Z_0),
  \end{split}
 \end{equation}
  where
  \begin{equation}
  \begin{split}
\ln(Z)-
\ln(Z_0)&=\int_0^1\mathrm{d}r\frac{1}{Z(r)}\frac{\partial}{\partial r}
Z(r)
  \end{split}
 \end{equation}
and $Z(r)=\tr{\mathrm{e}^{-\beta H_r}}$, $H_r=H_0+r(H-H_0)$. Now, we use the formula (see, e.g., section $6.5$ of \cite{Bhatia07})
 \begin{equation}
 \frac{\partial}{\partial r}\mathrm{e}^{-\beta H_r}=-\beta\int_0^1 \textrm{d}s\, \mathrm{e}^{-\beta sH_r}\left(H-H_0\right)\mathrm{e}^{-\beta (1-s) H_r}
\end{equation}
such that by the cyclic property of the trace
 \begin{equation}
\frac{\partial}{\partial r}
Z(r)= \beta \tr{\left(H_0-H\right)\mathrm{e}^{-\beta H_r}}
\end{equation}
and hence
   \begin{equation}
  \begin{split}
S(\langle\rho\rangle\|\rho_\beta)&\le 2\beta\|H-H_0\|.
  \end{split}
 \end{equation}

\section{Comparison with Robinson's construction}
\label{sec:Robinson}
In this section, we discuss Robinson's construction in Ref.~\cite{Rob73}. There, infinite lattice analogues of our results may be found and the key assumption in \cite{Rob73} is asymptotic abelianness, which effectively guarantees transport.  So a natural question to ask is whether there is a finite-size analogue of this assumption leading to similar behaviour.

Suppose we have a state $\omega$ that commutes with the Hamiltonian $H$.  And let $U_S$ be a unitary localized on a subsystem $S$.  Then suppose that at $t=0$ we apply $U_S$ to the state, getting $U_S \omega U_S^{\dagger}$.  This evolves over time as $e^{-\mathrm{i}Ht}(U_S \omega U_S^{\dagger})e^{\mathrm{i}Ht}$.

We assume that equilibration occurs to the time average state $\langle U_S\omega U_S^{\dagger}\rangle$.  Let $A_S$ be an observable on $S$, then the difference between the expectation values is
\begin{equation}
 \begin{split}
\tr{A_S\left(\omega - \langle U_S\omega U_S^{\dagger}\rangle\right)}
& = \tr{\langle A_S\rangle\left(\omega - U_S\omega U_S^{\dagger}\right)}\\
 & = \tr{\left(\langle A_S\rangle-U_S^{\dagger}\langle A_S\rangle U_S\right) \omega}\\
 & \leq \|\langle A_S\rangle-U_S^{\dagger}\langle A_S\rangle U_S\|\\
 & = \|\left[U_S,\langle A_S\rangle\right]\|,
 \end{split}
\end{equation}
where $\langle A_S\rangle=\lim_{T\rightarrow\infty}\f{1}{T}\int_0^T\mathrm{d} t\,e^{\mathrm{i}Ht}A_Se^{-\mathrm{i}Ht}$ is the time-average observable in the Heisenberg picture.  If we assume that the dynamics spreads $A_S(t)$ out over time, so that
\begin{equation}\label{eq:1}
 \|\left[U_S,\langle A_S\rangle\right]\|\rightarrow 0\ \textrm{as}\ N\rightarrow\ \infty,
\end{equation}
then the expectation values coincide as $N\rightarrow \infty$.  The assumption of asymptotic abelianness in \cite{Rob73} is a little different.  Because the setting is an infinite lattice, one can take limits of expectation values as time goes to infinity.  So the condition in \cite{Rob73} is essentially
\begin{equation}
 \|\left[U_S, A_S(t)\right]\|\rightarrow 0\ \textrm{as}\ t\rightarrow\ \infty.
\end{equation}
There are other technical assumptions that need to be made in the infinite lattice setting, but they are not important here.
It is not clear when one can verify that the condition in Eq.~(\ref{eq:1}) holds, except for simple cases.  Take the example of a translationally-invariant non-interacting free fermion model with non-degenerate {\it single-particle} energies.  Then taking an observable like $A_S = a^{\dagger}_na_n$, which counts the number of particles on site $n$, $\av{A_S} =\f{1}{N}\sum_n a^{\dagger}_na_n$.  Therefore, $\|\left[U_S,\langle A_S\rangle\right]\| = O(1/N)$.  This scaling is probably the best case scenario.  More generally, one probably gets slower decay with $N$ when the condition holds.


\begin{thebibliography}{10}
 \bibitem{EFG14}
J.~Eisert, M.~Friesdorf, and C.~Gogolin.
\newblock {Quantum many-body systems out of equilibrium}.
\newblock {\em Nature Physics}, 11:124, 2015.

\bibitem{GHRRS15}
J.~Goold, M.~Huber, A.~Riera, L.~del Rio, and P.~Skrzypczyk.
\newblock The role of quantum information in thermodynamics—a topical review.
\newblock {\em Journal of Physics A: Mathematical and Theoretical},
  49(14):143001, 2016.

\bibitem{GE15}
C.~Gogolin and J.~Eisert.
\newblock {Equilibration, thermalisation and the emergence of statistical
  mechanics in closed quantum systems}.
\newblock {\em Rep. Prog. Phys.}, 79:056001, 2016.

\bibitem{Reimann10}
P.~Reimann.
\newblock Canonical thermalization.
\newblock {\em New Journal of Physics}, 12(5):055027, 2010.

\bibitem{RGE11}
A.~Riera, C.~Gogolin, and J.~Eisert.
\newblock Thermalization in nature and on a quantum computer.
\newblock {\em Phys. Rev. Lett.}, 108:080402, 2012.

\bibitem{MAMW13}
M.~P. M{\"u}ller, E.~Adlam, L.~Masanes, and N.~Wiebe.
\newblock Thermalization and canonical typicality in translation-invariant
  quantum lattice systems.
\newblock {\em Comm. Math. Phys.}, 340(2):499--561, 2015.

\bibitem{BC15}
F.~G. S.~L. Brand\~ao and M.~Cramer.
\newblock {Equivalence of Statistical Mechanical Ensembles for Non-Critical
  Quantum Systems}.
\newblock arXiv:1502.03263, 2015.

\bibitem{Tasaki16}
H.~Tasaki.
\newblock {On the local equivalence between the canonical and the
  microcanonical distributions for quantum spin systems}.
\newblock {arXiv:1609.06983}, 2016.

\bibitem{Tasaki98}
H.~Tasaki.
\newblock From quantum dynamics to the canonical distribution:\ general picture
  and a rigorous example.
\newblock {\em Phys. Rev. Lett.}, 80:1373, 1998.

\bibitem{LPSW09}
N.~Linden, S.~Popescu, A.~J. Short, and A.~Winter.
\newblock Quantum mechanical evolution towards thermal equilibrium.
\newblock {\em Phys. Rev. E}, 79:061103, 2009.

\bibitem{Reimann08}
P.~Reimann.
\newblock Foundation of statistical mechanics under experimentally realistic
  conditions.
\newblock {\em Phys. Rev. Lett.}, 101:190403, 2008.

\bibitem{SPTD15}
J.~Schachenmayer, L.~Pollet, M.~Troyer, and A.~J. Daley.
\newblock Thermalization of strongly interacting bosons after spontaneous
  emissions in optical lattices.
\newblock {\em EPJ Quantum Technology}, 2(1):1--14, 2015.

\bibitem{Rob73}
D.~W. Robinson.
\newblock Return to equilibrium.
\newblock {\em Comm. Math. Phys.}, 31(3):171--189, 1973.

\bibitem{HR86}
L.~Hume and D.~W. Robinson.
\newblock Return to equilibrium in the {XY} model.
\newblock {\em Journal of Statistical Physics}, 44(5):829--848, 1986.


\bibitem{Paulsen02}
V.~Paulsen.
\newblock {\em Completely Bounded Maps and Operator Algebras}.
\newblock Cambridge University Press, Cambridge, 2002.

\bibitem{AL70}
H.~Araki and E.~H. Lieb.
\newblock Entropy inequalities.
\newblock {\em Comm. Math. Phys.}, 18:160–170, 1970.

\bibitem{Bhatia07}
R.~Bhatia.
\newblock {\em Positive Definite Matrices}.
\newblock Princeton University Press, New Jersey, 2007.

\bibitem{BFS00}
V.~Bach, J.~Fr{\"o}hlich, and I.~M. Sigal.
\newblock Return to equilibrium.
\newblock {\em Journal of Mathematical Physics}, 41(6):3985--4060, 2000.

\bibitem{JPPP15}
V.~Jak{\v{s}}i{\'{c}}, J.~Panangaden, A.~Panati, and C.~Pillet.
\newblock Energy conservation, counting statistics, and return to equilibrium.
\newblock {\em Letters in Mathematical Physics}, 105(7):917--938, 2015.

\bibitem{HF16}
W.~Hahn and B.~V. Fine.
\newblock {Stability of Quantum Statistical Ensembles with Respect to Local
  Measurements}.
\newblock arXiv:1601.06402, 2016.

\bibitem{KGKRE14}
M.~Kliesch, C.~Gogolin, M.~J. Kastoryano, A.~Riera, and J.~Eisert.
\newblock Locality of temperature.
\newblock {\em Phys. Rev. X}, 4:031019, 2014.

\bibitem{HK06}
M.~B. Hastings and T.~Koma.
\newblock Spectral gap and exponential decay of correlations.
\newblock {\em Comm. Math. Phys.}, 265(3):781--804, 2006.

\bibitem{SF12}
A.~J. Short and T.~C. Farrelly.
\newblock Quantum equilibration in finite time.
\newblock {\em New Journal of Physics}, 14(1):013063, 2012.

\bibitem{CBG16}
M.~Cramer, F.~G. S.~L. Brand\~ao, and M.~Guta.
\newblock A {B}erry-{E}sseen theorem for quantum lattice systems.
\newblock {\em In preparation}.

\bibitem{Short10}
A.~J. Short.
\newblock Equilibration of quantum systems and subsystems.
\newblock {\em New Journal of Physics}, 13(5):053009, 2011.

\bibitem{BCH11}
M.~C. Ba\~nuls, J.~I. Cirac, and M.~B. Hastings.
\newblock Strong and weak thermalization of infinite nonintegrable quantum
  systems.
\newblock {\em Phys. Rev. Lett.}, 106:050405, 2011.

\bibitem{GHT13}
S.~Goldstein, T.~Hara, and H.~Tasaki.
\newblock Time scales in the approach to equilibrium of macroscopic quantum
  systems.
\newblock {\em Phys. Rev. Lett.}, 111:140401, 2013.

\bibitem{MGLFS14}
A.~S.~L. Malabarba, L.~P. Garc\'{i}a-Pintos, N.~Linden, T.~C. Farrelly, and
  A.~J. Short.
\newblock Quantum systems equilibrate rapidly for most observables.
\newblock {\em Phys. Rev. E}, 90:012121, 2014.

\bibitem{CDEO08}
M.~Cramer, C.~M. Dawson, J.~Eisert, and T.~J. Osborne.
\newblock Exact relaxation in a class of nonequilibrium quantum lattice
  systems.
\newblock {\em Phys. Rev. Lett.}, 100:030602, 2008.

\bibitem{TCF15}
T.~C. Farrelly.
\newblock Equilibration of quantum gases.
\newblock {\em New Journal of Physics}, 18(7):073014, 2016.

\bibitem{VZ12}
Vinayak and M.~Žnidarič.
\newblock Subsystem dynamics under random {H}amiltonian evolution.
\newblock {\em Journal of Physics A: Mathematical and Theoretical},
  45(12):125204, 2012.

\bibitem{Cramer12}
M.~Cramer.
\newblock Thermalization under randomized local {H}amiltonians.
\newblock {\em New Journal of Physics}, 14(5):053051, 2012.

\bibitem{BCHHKM12}
F.~G. S.~L. Brand\~ao, P.~\ifmmode \acute{C}\else
  \'{C}\fi{}wikli\ifmmode~\acute{n}\else \'{n}\fi{}ski, M.~Horodecki,
  P.~Horodecki, J.~K. Korbicz, and M.~Mozrzymas.
\newblock Convergence to equilibrium under a random {H}amiltonian.
\newblock {\em Phys. Rev. E}, 86:031101, 2012.

\bibitem{UWE12}
C.~Ududec, N.~Wiebe, and J.~Emerson.
\newblock Information-theoretic equilibration:\ the appearance of
  irreversibility under complex quantum dynamics.
\newblock {\em Phys. Rev. Lett.}, 111:080403, 2013.

\bibitem{MRA13}
L.~Masanes, A.~Roncaglia, and A.~Ac\'{i}n.
\newblock Complexity of energy eigenstates as a mechanism for equilibration.
\newblock {\em Phys. Rev. E}, 87:032137, 2013.

\bibitem{Reimann16}
P.~Reimann.
\newblock Typical fast thermalization processes in closed many-body systems.
\newblock {\em Nature Communications}, 7:10821, 2016.

\bibitem{SSM15}
M.~Schiulaz, A.~Silva, and M.~M{\"u}ller.
\newblock Dynamics in many-body localized quantum systems without disorder.
\newblock {\em Phys. Rev. B}, 91:184202, 2015.

\bibitem{Araki69}
H.~Araki.
\newblock Gibbs states of a one dimensional quantum lattice.
\newblock {\em Comm. Math. Phys.}, 14(2):120--157, 1969.






\end{thebibliography}

\end{document}